\RequirePackage{amssymb, amsthm, amsmath, mathtools}

\documentclass[12pt]{iopart}

\usepackage[square,sort&compress,comma,numbers]{natbib}

\usepackage{hyperref}

\newcommand{\MyCases}[4]{%
\left\{\begin{array}{l@{\quad}l}%
#1 & #2 \\%
#3 & #4%
\end{array}\right.%
}

\newcommand{\changed}[2]{#1} % new old

\newtheorem{definition}{Definition}

\newtheorem{example}{Example}
\newtheorem{theorem}{Theorem}
\newtheorem{problem}{Problem}
\newtheorem{remark}{Remark}

\usepackage{algorithm}
\usepackage[noend]{algpseudocode}

\usepackage{subfloat}
\newcommand{\De}{\coloneqq}

\def\Diag{\operatorname{diag}}
\def\Prox{\operatorname{prox}}

\def\bR{\mathbb R}
\def\bRI{\bR_\infty}

\newcommand{\ie}{\textit{i.e.}}
\newcommand{\eg}{\textit{e.g.}}

\newcommand{\dom}{\operatorname{dom}}
\newcommand{\TV}{\operatorname{TV}}
\newcommand{\dTV}{\operatorname{dTV}}
\newcommand{\TGV}{\operatorname{TGV}}

\def\mA{\mathbf A}

\def\mD{\mathbf D}
\def\mI{\mathbf I}
\def\mP{\mathbf P}
\def\mS{\mathbf S}
\def\mT{\mathbf T}

\def\dataFdg{\texttt{FDG}}
\def\dataAmy{\texttt{florbetapir}}

\usepackage{tikz}

\newcommand{\PlotAmyX}[4]{% left, bottom, right, top
\includegraphics[clip, trim=40px 0px 42px 30px, height=#4]{#1_amy_#2_x85_#3}}
\newcommand{\PlotAmyY}[4]{% left, bottom, right, top
\includegraphics[clip, trim=12px 0px 38px 30px, height=#4]{#1_amy_#2_y85_#3}}
\newcommand{\PlotAmyZ}[4]{% left, bottom, right, top
\includegraphics[clip, trim=43px 45px 43px 30px, height=#4]{#1_amy_#2_z63_#3}}

\newcommand{\PlotFdgX}[4]{% left, bottom, right, top
\includegraphics[clip, trim=45px 40px 47px 0px, height=#4]{#1_fdg_#2_x85_#3}}
\newcommand{\PlotFdgY}[4]{% left, bottom, right, top
\includegraphics[clip, trim=32px 40px 48px 10px, height=#4]{#1_fdg_#2_y90_#3}}
\newcommand{\PlotFdgZ}[4]{% left, bottom, right, top
\includegraphics[clip, trim=43px 45px 43px 30px, height=#4]{#1_fdg_#2_z46_#3}}

\usepackage{transparent}

\newcommand{\DrawLabel}[1]{%
\node[anchor=south west, xshift=1pt, yshift=1pt] at (0,0) {{\transparent{0.5}\colorbox{black}{\color{white}\textbf{\transparent{1}\phantom{fg}\hspace{-2mm}#1}}}};}%

\def\PicWidth{3.5cm}

\newcommand{\PlotAmy}[4]{%
\begin{tikzpicture}%
\node[anchor=south west] at (0,0) {\PlotAmyX{#2}{#3}{#4}{\PicWidth}\PlotAmyY{#2}{#3}{#4}{\PicWidth}\PlotAmyZ{#2}{#3}{#4}{\PicWidth}};%
\DrawLabel{#1}%
\end{tikzpicture}%
}%

\newcommand{\PlotFdg}[4]{%
\begin{tikzpicture}%
\node[anchor=south west] at (0,0) {\PlotFdgX{#2}{#3}{#4}{\PicWidth}\PlotFdgY{#2}{#3}{#4}{\PicWidth}\PlotFdgZ{#2}{#3}{#4}{\PicWidth}};%
\DrawLabel{#1}%
\end{tikzpicture}%
}%

\newcommand{\PlotAmyZY}[4]{%
\begin{tikzpicture}%
\node[anchor=south west] at (0,0) {\PlotAmyZ{#2}{#3}{#4}{\PicWidth}\PlotAmyY{#2}{#3}{#4}{\PicWidth}};%
\DrawLabel{#1}%
\end{tikzpicture}%
}%

\newcommand{\PlotFdgZY}[4]{%
\begin{tikzpicture}%
\node[anchor=south west] at (0,0) {\PlotFdgZ{#2}{#3}{#4}{\PicWidth}\PlotFdgY{#2}{#3}{#4}{\PicWidth}};%
\DrawLabel{#1}%
\end{tikzpicture}%
}%

\newcommand{\InclFig}[1]{\centering\includegraphics[clip, trim=7pt 7pt 7pt 17pt, width=\linewidth]{#1}%
}%

\newcommand{\InclFigSmall}[1]{\centering\includegraphics[clip, trim=7pt 7pt 7pt 7pt, width=0.71\linewidth]{#1}%
}%

\begin{document}
\title[Faster PET Reconstruction by Randomization and Preconditioning]{Faster PET Reconstruction with Non-Smooth Priors by Randomization and Preconditioning}

\author{Matthias~J.~Ehrhardt$^1$, Pawel~Markiewicz$^2$ \& Carola-Bibiane Sch\"onlieb$^3$}

\address{$^1$ Institute for Mathematical Innovation, University of Bath, Bath BA2 7JU, UK \\ $^2$ Centre for Medical Image Computing, London WC1E 6BT, UK \\ $^3$ Department for Applied Mathematics and Theoretical Physics, University of Cambridge, Cambridge CB3 0WA, UK}
\ead{m.ehrhardt@bath.ac.uk}

\vspace{10pt}
\begin{indented}
\item[]July 2019
\end{indented}

\begin{abstract}
Uncompressed clinical data from modern positron emission tomography (PET) scanners are very large, exceeding 350 million data points (projection bins). The last decades have seen tremendous advancements in mathematical imaging tools many of which lead to non-smooth (\ie~non-differentiable) optimization problems which are much harder to solve than smooth optimization problems. Most of these tools have not been translated to clinical PET data, as the state-of-the-art algorithms for non-smooth problems do not scale well to large data. In this work, inspired by big data machine learning applications, we use advanced randomized optimization algorithms to solve the PET reconstruction problem for a very large class of non-smooth priors which includes for example total variation, total generalized variation, directional total variation and various different physical constraints. The proposed algorithm randomly uses subsets of the data and only updates the variables associated with these. While this idea often leads to divergent algorithms, we show that the proposed algorithm does indeed converge for any proper subset selection. Numerically, we show on real PET data (FDG and florbetapir) from a Siemens Biograph mMR that about ten projections and backprojections are sufficient to solve the MAP optimisation problem related to many popular non-smooth priors; thus showing that the proposed algorithm is fast enough to bring these models into routine clinical practice.
\end{abstract}

\vspace{2pc}

%\submitto{}
\newpage

\section{Introduction}
Positron emission tomography (PET) is an important clinical imaging technique as it allows monitoring function of the human body by following a radio-active tracer. The image reconstruction process in PET is challenging as the low number of photon counts call for the Poisson noise modeling and the amount of data is excessively large on modern scanners. While most clinical systems still run non-penalized reconstructions, it has been shown that priors can improve noise control and quantification~\cite{Teoh2015a, Ahn2015}. In addition, the research of the last decade suggests that non-smooth priors, such as the total variation~\cite{Rudin1992ROF} and its relatives like total generalized variation~\cite{Benning2010, Bredies2010, Bredies2015tgvnumerics}, are beneficial for imaging applications as they allow smooth variations within regions without oversmoothing sharp boundaries~\cite{Rudin1992ROF, Setzer2010, Benning2010, Bredies2010, Anthoine2011a, Burger2013, Sawatzky2013,Bredies2015tgvnumerics, Zhang2016, Ehrhardt2016b}. These priors have been widely studied in the context of PET (\eg~\cite{Sawatzky2008, Guo2009, Ahn2012, Muller2012, Cabello2013, ChenyeWang2014, Wang2015}) and other medical imaging modalities, \eg~computed tomography (CT) \cite{Niu2014, Gu2018}, photoacoustic tomography (PAT) \cite{Boink2018}, magnetic resonance imaging (MRI) \cite{Knoll2011, Ehrhardt2016b}. Modern PET scanners always come with a second anatomical modality such as CT or MRI. Non-smooth priors can also be used to either incorporate anatomical knowledge from MRI or CT into the reconstruction,~\eg~\cite{Bowsher2004, Hintermuller2017, Ehrhardt2016a, Ehrhardt2016b, Schramm2017petplusmri, Mehranian2017petplusmri}, or to jointly reconstruct PET and the anatomical CT/MRI image~\cite{Ehrhardt2015petmri, Knoll2016, Rasch2017, Mehranian2017petmri}. Only a few optimization algorithms are capable of combining non-smooth priors and the Poisson noise model, \eg~\cite{Setzer2010, Esser2010, Chambolle2011, Pock2011, Dupe2011, Figueiredo2010, Krol2012, Harmany2012, Sawatzky2013, Wang2015, Lin2019} and most of these are not applicable to solve all regularization models mentioned above.

One of the most popular algorithms to solve the resulting non-smooth convex optimization problem is the primal-dual hybrid gradient (PDHG) algorithm\footnote{also known as the "Chambolle--Pock algorithm"}~\cite{Esser2010, Chambolle2011, Pock2011}. PDHG has been used in numerous imaging studies on multiple imaging modalities, including PET, see \eg~\cite{Dupe2011, Wolf2013, Rigie2015tnvct, Zhang2016, FoygelBarber2016, Knoll2016, Schramm2017petplusmri, Rasch2017, Rigie2017}. While this algorithm is flexible enough to solve a variety of non-smooth optimization problems, in every iteration both the projection and the backprojection have to be applied for all projection bins. Moreover, in every iteration computations on vectors that have the size of the data have to be performed. For modern scanners like the Siemens Biograph mMR with span-1 data format, these vectors contain more than 350 million elements and therefore limiting the applicability of this algorithm (and thus many non-smooth priors) to state-of-the-art scanners.

\subsection{Contributions}

\textbf{Subset Acceleration with Randomization} We propose an algorithm, coined Stochastic PDHG or SPDHG for short, which in every iteration performs computations only for a random subset of the data. We show on clinical data from a Siemens Biograph mMR that with this algorithm, for the first time, non-smooth priors become feasible to be used in routine clinical imaging.  Numerically, we show that SPDHG is competitive with OSEM on unregularized reconstruction problems but stable with respect to the choice of the subsets due to its mathematically guaranteed convergence. In fact, SPDHG converges to the deterministic solution for any proper subset selection, see Theorem \ref{THE:ALG}.

In addition to the general randomized solution strategy, we propose two further algorithmic advancements: preconditioning and non-uniform sampling.

\textbf{Preconditioning} We propose and evaluate the use of data-dependent preconditioners in SPDHG for PET image reconstruction. While the convergence theory for a large class of preconditioners has been available since 2011~\cite{Pock2011}, our proposed preconditioners are the first to be computationally efficient and effective for PET image reconstruction with non-smooth priors. The speed enhancement of preconditioning for PDHG was recognized before~\cite{Sidky2012}, however, we present a novel formulation of these preconditioners that is computiationally efficient, see Theorem~\ref{THE:STEPPARM}.

\textbf{Non-uniform Sampling} We propose a novel non-uniform sampling strategy, which is necessary to accommodate the differences of data fidelity and regularity. Both randomization and preconditioning can be used independently or can be combined as proposed here in this work.

\textbf{Uncompressed Data} In this work we use uncompressed (span-1) data from the Siemens Biograph mMR. While it is not clear if and how much this improves the reconstructed PET images~\cite{Belzunce2017}, the proposed algorithm is fast enough to study the benefits of uncompressed data in combination with a variety of regularization models.

A few initial findings on randomized reconstruction without preconditioning were published in a conference paper~\cite{Ehrhardt2017SPIE}.

\subsection{PET Reconstruction via Optimization}
Given the measured data vector $b \in \mathbb N^M$ and the projection model $\mP$, the PET reconstruction problem can be formulated as the solution to the optimization problem
\begin{align}
    \min_{u \geq 0} \Bigl\{D(\mP u) + \alpha R(u) \Bigr\} \label{EQU:PROB:OPT}
\end{align}
where the data fidelity $D(\mP u)$ measures the match of the estimated image $u$ with the data and the prior $\alpha R(u)$ penalizes features that are not desirable in the solution. In other words the prior can be used to avoid solutions which would fit the noisy data too closely. The data fidelity $D$ is (up to constants independent of $u$) the negative log-likelihood of the multi-variate Poisson distribution
\begin{align*}
  D(y) = \sum_{i=1}^M y_i + r_i - b_i + b_i \log\left( \frac{b_i}{y_i + r_i}\right) \, ,
\end{align*}
with expected value being the sum of the projected image $y$ and the estimated background activity $r$. The latter is needed in order to model non-linear effects such as scatter and randoms. The data fidelity $D$ measures the distance of the estimated data $\mP u + r$ to the measured data $b$ in the sense that $D(\mP u) \geq 0$ and $D(\mP u) = 0$ if and only if $\mP u + r = b$. The operator $\mP$ performs the projection and includes geometric factors, attenuation and normalization.

While the main motivation is the efficient solution of non-smooth optimization problems, we first compare the method to \textit{ordered subsets expectation maximization} (OSEM)~\cite{Hudson1994} for unregularized reconstruction. The "ordered subsets" idea has subsequently been used for many algorithms related to non-smooth optimisation, see \eg~\cite{McGaffin2015}. We would like to show in the next example 1) that the "ordered subsets" idea is generally non-convergent and thus may be unstable and 2) that the proposed algorithm is as fast as OSEM---despite its proven convergence.

\subsection{Motivating Example: OSEM}
\begin{subfigures}%
\begin{figure}%
\InclFigSmall{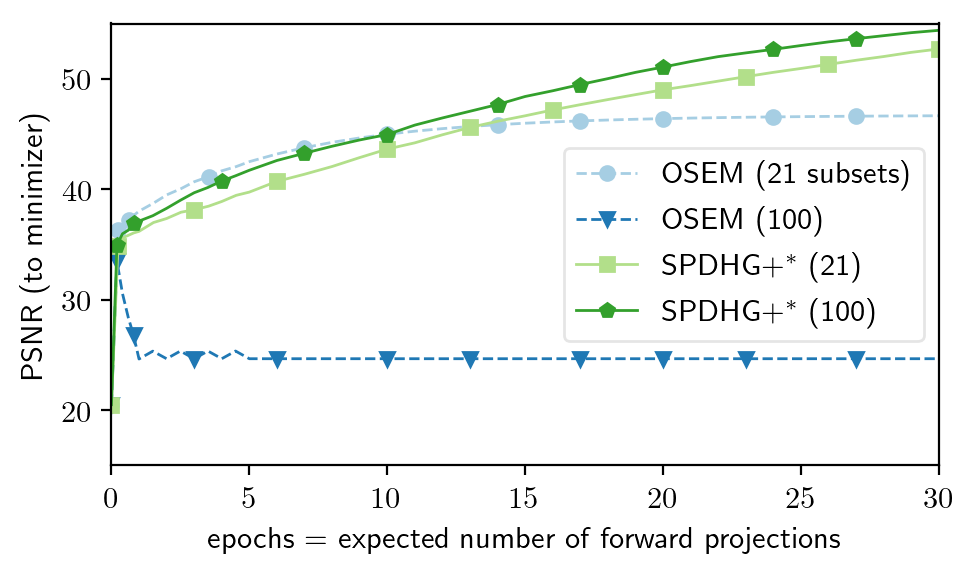}%
\caption{\textbf{OSEM may become unstable.} OSEM and SPDHG+ are compared for a varying number of subsets. While the speed of SPDHG+ increases with the number of subsets, OSEM fails to converge to the right solution for 100 subsets. $^\ast$proposed} \label{FIG:STABILITY:QUANT}%
\end{figure}%
\begin{figure}%
\centering%
\PlotAmy{OSEM}{ml_bin}{inferno}{smoothed_OSEM-21_50}\\%
\PlotAmy{SPDHG+ (proposed)}{ml_bin}{inferno}{smoothed_SPDHG2-21_50}%
\caption{\textbf{OSEM may become unstable II.} In this example both OSEM and SPDHG+ take 21 \textbf{subsets with bins equidistantly divided} into 21 subsets. In contrast to OSEM, SPDHG+ is robust with respect to this subset selection and achieves a reasonable solution.} \label{FIG:STABILITY:QUAL}%
\end{figure}%
\begin{figure}%
\InclFigSmall{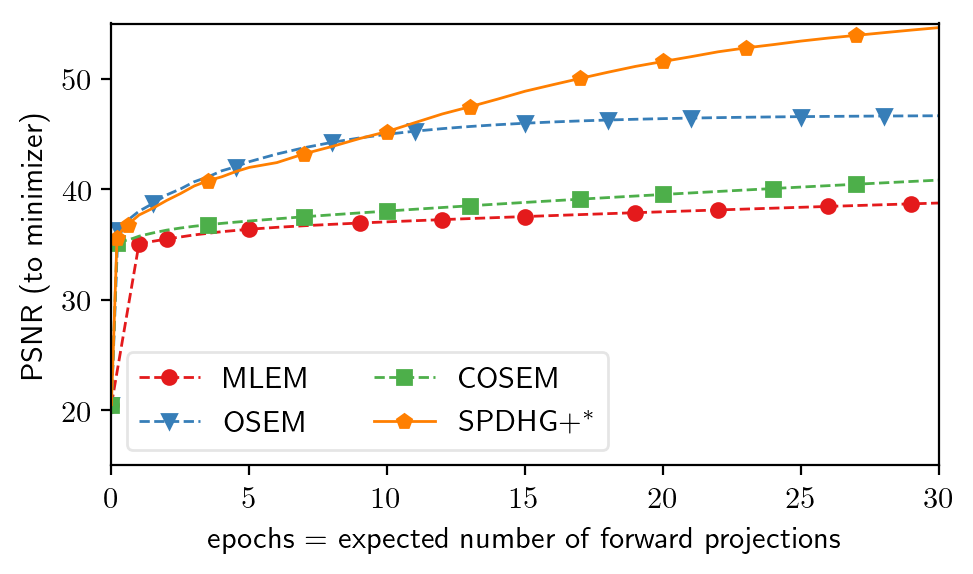}%
\caption{\textbf{Faster with subsets.} Comparison of reconstruction speed of several algorithms. We compare MLEM, OSEM (21~subsets), COSEM (252~subsets) and the proposed SPDHG+ (252~subsets) in terms of $\operatorname{PSNR}(x^k, x^\ast)$ (see section \ref{sec:numerics}) where $x^\ast$ is an optimal solution for the \dataAmy~dataset (see section \ref{section:data}) approximated by 5k MLEM iterations. The subsets are selected with angles equidistantly divided. OSEM and SPDHG+ are clearly faster than MLEM and COSEM. $^\ast$proposed} \label{FIG:MLEM:OSEM} %
\end{figure}%
\begin{figure}
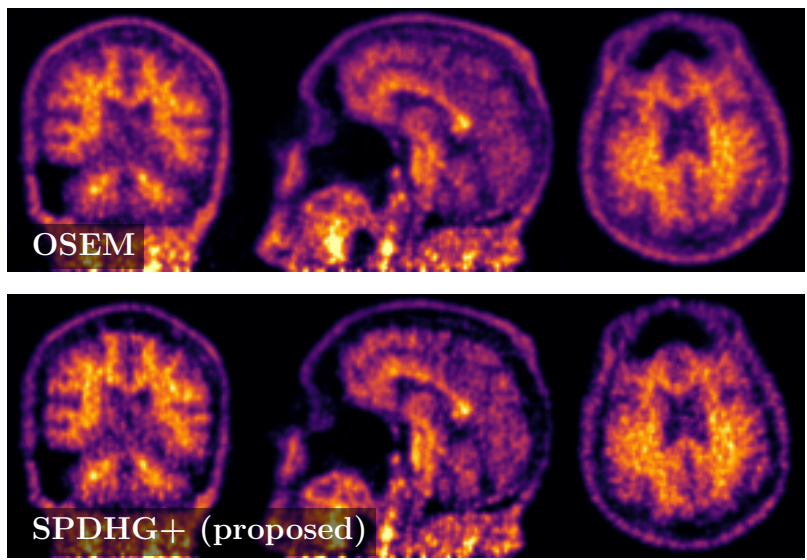
%
\centering%
\PlotAmy{OSEM}{ml}{inferno}{smoothed_OSEM-21_10}\\%
\PlotAmy{SPDHG+ (proposed)}{ml}{inferno}{smoothed_SPDHG2-252_10}%
\caption{\textbf{OSEM and SPDHG+ look the same.} Visual comparison of OSEM (21~subsets) and SPDHG+ (252) after 10 epochs for maximum likelihood reconstruction. Both algorithms achieve very similar images.} \label{FIG:OSEM:IMAGES}%
\end{figure}%
\end{subfigures}%

If there is no prior, \ie~$\alpha R = 0$, the most common algorithm to solve the optimization problem~\eqref{EQU:PROB:OPT} is the \textit{maximum likelihood expectation maximization} algorithm (MLEM)~\cite{Shepp1982} defined by
\begin{align}
  u^{k+1} = \frac{u^k}{\mP^T 1} \mP^T\left(\frac{b}{\mP u^k + r} \right) \label{EQU:MLEM} \, ,
\end{align}
where all operations have to be understood element-wise. The computational bottleneck in the MLEM algorithm is the evaluation of the operator $\mP$ and its transpose $\mP^T$ in each iteration.

To overcome this hurdle, it has been proposed to change the update and evaluate the operator and its adjoint only on one out of $m$ subsets of the data in each iteration. At every iteration $k$ we choose $i = \operatorname{mod}(k, m)$ and change update formula~\eqref{EQU:MLEM} to
\begin{align}
  u^{k+1} = \frac{u^k}{\mP^T_i 1} \mP^T_i\left( \frac{b_i}{\mP_i u^k + r_i} \right) \label{EQU:OSEM}  \, .
\end{align}
This algorithm became known as OSEM. Here $\mP_i$ is the restriction of $\mP$ onto the $i$th subset, \ie~$\mP = (\mP_1^T, \ldots, \mP_m^T)^T$. While this change of the update equation reduces the computational burden by $1 / m$, it is in general not guaranteed to converge to a solution of~\eqref{EQU:PROB:OPT}, illustrated in Figures~\ref{FIG:STABILITY:QUANT} and~\ref{FIG:STABILITY:QUAL}. A convergent version of OSEM, called \textit{complete-data OSEM} (COSEM), has been developed~\cite{Hsiao2002cosem}. While it comes with mathematical convergence guarantees, it is much slower than OSEM (see Figure~\ref{FIG:MLEM:OSEM}) and therefore never became popular for the reconstruction of clinical PET data.

MLEM has been extended to include smooth~\cite{Green1990} and certain non-smooth~\cite{Sawatzky2013} prior information, however, conceptually both algorithms intrinsically struggle with the ordered subset acceleration. Also other algorithms have been ``accelerated'' based on the ordered subset idea, \eg~\cite{McGaffin2015, RossSchmidtlein2017}, but are similarly intrinsically unstable due to their non-convergence. See~\cite{Cheng2013} for a numerical comparison and~\cite{Teoh2015a, Ahn2015} for a validation on clinical PET data. For differentiable priors, a surrogate based technique allows for stable subset acceleration~\cite{DePierro2001, Ahn2003, Ahn2006}. In this work we propose the subset-accelerated algorithm SPDHG that is provably convergent and thus stable and robust, see Figures~\ref{FIG:STABILITY:QUANT} and~\ref{FIG:STABILITY:QUAL}. SPDHG is flexible enough to be applicable to a large variety of convex and non-smooth priors and is as efficient as OSEM if no explicit prior is being used, see Figures~\ref{FIG:MLEM:OSEM} and~\ref{FIG:OSEM:IMAGES}.

\section{Mathematical Model}%
\subsection{Non-Smooth PET Reconstruction with Subsets}%
As outlined above, PET reconstruction can be formulated in terms of the optimization problem~\eqref{EQU:PROB:OPT}. Computationally, it is convenient to rewrite (and solve) the optimization problem~\eqref{EQU:PROB:OPT} in terms of subsets. We denote by $M$ the number of projection bins. Let $\{S_i\}$ be a partition of $[M]$, in the sense that $\cup_{i=1}^m S_i = [M]$, where we used the notation $[M] \De \{1, \ldots, M\}$. It is not necessary to assume that $S_i \cap S_j = \emptyset$ for $i \neq j$. For notational simplicity we will restrict ourselves to the this case. We define
\begin{align}
    D_i(y) \De \sum_{j \in S_i} \varphi_j(y_j) \label{EQU:DATATERM}
\end{align}
with the distance function for every data point given by
\begin{align}
    \varphi(y) \De \MyCases{y + r - b \log(y + r) - b + b \log b }{\text{if $y + r \geq 0$}}%
    {\infty}{\text{else}} \, ,\label{EQU:LOCALDISTANCE}
\end{align}
where we omitted the index $j$ at $\varphi, y, r$ and $b$ for readability. Algorithms from convex optimization require the problem to be defined over an entire vector space which we satisfy by extending $\varphi$ to $\infty$ for non-positive estimated data $y + r$. The data and the background are photon counts and therefore have a natural non-negativity constraint. To allow for the concise notation in~\eqref{EQU:LOCALDISTANCE}, we define $0 \log 0 \De 0$ and $- \log 0 \De \infty$.

We model the non-negativity constraint for the image $u$ with the indicator function $\imath_+$, which is defined as
\begin{align}
  \imath_+(u) = \MyCases{0}{\text{if $u \geq 0$}}{\infty}{\text{else}} \, . \label{EQU:NONNEGATIVITY}
\end{align}
Thus, this results in the unconstrained optimization problem
\begin{problem}[PET Reconstruction with Subsets] \label{PROB:PET:SUBSETS}%
\begin{align}
   u^\sharp \in \arg \min_{u \in \bR^N} \left\{ \sum_{i=1}^m D_i(\mP_i u) + \alpha R(u) + \imath_+(u) \right\} \label{EQU:PROB:OPT:SUBSETS} \, .
\end{align}
\end{problem}
We would like to stress that solving problem~\eqref{EQU:PROB:OPT:SUBSETS} is equivalent to solving the original problem~\eqref{EQU:PROB:OPT} for any choice of subsets. In fact, the subset selection becomes a reconstruction parameter that may be varied to speed up the reconstruction procedure.

Often, our prior assumptions involve linear operators, too. One of the most prominent examples of this is the total variation~\cite{Rudin1992ROF} $$R(u) = \TV(u) = \|\nabla u\|_{2,1} = \sum_i \|\nabla u_i\|_2 = \sum_i \left(\sum_{j=1}^3 (\partial_j u_i)^2\right)^{1/2} \, ,$$ where we take the 2-norm locally, \ie~at every voxel $i$ we take the 2-norm of the spatial gradient, and the 1-norm globally, \ie~we sum over all voxels. Forward difference discretization of the gradient operator $\nabla$ is used as in~\cite{Chambolle2011}. Similarly, we use the directional total variation $R(u) = \dTV(u) = \|\mD \nabla u\|_{2,1}$ to incorporate a-priori knowledge about the solution given by an anatomical prior image, see~\cite{Ehrhardt2016a, Ehrhardt2016b, Schramm2017petplusmri, Bungert2018remotesensing} for details.

Solving problem~\eqref{EQU:PROB:OPT:SUBSETS} is challenging, even when the involved variables are small and matrix-vector products are easy to compute. The difficulty stems from its non-smoothness. The data term $D_i$ is not finite everywhere and while it is differentiable on its effective domain $\dom(D_i) := \{y \mid D_i(y) < \infty\}$, the gradient is not globally Lipschitz continuous. In addition, further non-smoothness comes from the constraint $\imath_+$ and the prior $R$ may be non-smooth as well. All of this being said, in PET reconstruction, the variable sizes are actually very large and matrix-vector products expensive to compute.

To apply optimization algorithms to solve~\eqref{EQU:PROB:OPT:SUBSETS}, we reformulate it as a generic optimization problem of the form
\begin{problem}[Generic Optimization Problem] \label{PROB:OPT:GENERIC}%
\begin{align}
   x^\sharp \in \arg \min_{x \in X} \left\{\Psi(x) := \sum_{i=1}^n f_i(\mA_i x) + g(x) \right\} \, . \label{EQU:PROB:OPT:GENERIC}
\end{align}
\end{problem}

For instance, for unregularized reconstructions, \ie~$\alpha R = 0$, we may make the association
\begin{align*}
    n = m, \quad g = \imath_+, \quad f_i &= D_i, \quad \mA_i = \mP_i
\end{align*}
and reconstructions regularized by the total variation, \ie~$R(u) = \|\nabla u\|_{2,1}$, can be achieved by
\begin{align}
\begin{aligned}
    n &= m + 1, & f_i &= D_i, i \in [m], & f_n &= \alpha \|\cdot\|_{2,1}\\
    g &= \imath_+, & \mA_i &= \mP_i, i \in [m], & \mA_n &= \nabla \, .
\end{aligned} \label{EQU:SETTING:TV}
\end{align}

\subsection{Optimization with Saddle-Point Problems} \label{SEC:OPTIMIZATION}%
Instead of solving problem~\eqref{EQU:PROB:OPT:GENERIC} directly, it is more efficient to reformulate the minimization problem as a saddle point problem making use of the \emph{convex conjugate} of a functional, see \eg~\cite{Bauschke2011}.

\begin{definition}[Convex Conjugate]
Let $f : Y \to \bRI \De \bR \cup \{\infty\}$ be a functional with extended real values. Then we define the convex conjugate of $f$ as $f^\ast : Y \to \bRI$ with
\begin{align*}
f^\ast(y) = \sup_x \left\{ \langle y, x\rangle - f(x) \right\} \, .
\end{align*}
\end{definition}

For convex, proper and lower semi-continuous (lsc) functionals $f$ we have that $f^{\ast\ast} = f$, see \eg~\cite{Bauschke2011}, and thus $f(x) = \sup_y \left\{ \langle x, y\rangle - f^\ast(y) \right\}$. Then, with $Y = \prod_{i=1}^n Y_i$, problem~\eqref{EQU:PROB:OPT:GENERIC} is equivalent to

\begin{problem}[Generic Saddle Point Problem] \label{PROB:SADDLE}%
\begin{align}
   \min_{x \in X} \sup_{y \in Y} \left\{ \sum_{i=1}^n \langle \mA_i x, y_i \rangle - f_i^\ast (y_i) + g(x) \right\} \, . \label{EQU:SADDLE}
\end{align}
\end{problem}
We will refer to the variable $x$ as the \emph{primal variable} and to $y$ as the \emph{dual variable}.

\begin{example}
The convex conjugate of the PET distance function \eqref{EQU:DATATERM} is given by $D_i^\ast(y) = \sum_{j \in S_i} \varphi^\ast_j(y_j)$ with
\begin{align}
     \varphi^\ast(y) = \MyCases{-y r - b \log(1 - y)}{\text{if $y \leq 1$}}{\infty}{\text{else}} \label{EQU:PET:CVX}
\end{align}
where we omitted the index $j$ at $\varphi, y, b$ and $r$ for readability.
\end{example}
The derivation of the formulas in this and the following example are omitted for brevity.

As some (or all) of the $f_i$ and $g$ in~\eqref{EQU:PROB:OPT:GENERIC} are non-smooth, we make use of the proximal operator of these. Our definition varies slightly from the usual definition as we allow the step size parameter to be matrix-valued. For a symmetric and positive definite matrix $\mS$, we define the weighted norm $\|x\|_\mS$ as $\|x\|_\mS^2 \De \|\mS^{-1/2} x\|^2 = \langle \mS^{-1} x, x \rangle$.

\begin{definition}[Proximal Operator]
Let $\mS$ be a symmetric and positive definite matrix. Then we define the proximal operator of $f$ with metric (or step size) $\mS$ as
\begin{align*}
\Prox^\mS_f(x)
&\De \arg \min_z \left\{ \|z - x\|^2_\mS + f(z) \right\} \, .
\end{align*}
\end{definition}
From here on, $\mS$ and $\mT$ will always be diagonal (and thus symmetric) and positive definite matrices.

\begin{example}
The proximity operator of the non-negativity constraint \eqref{EQU:NONNEGATIVITY} is given element-wise by
\begin{align*}
    \Prox^\mT_{\imath_+}(x) = \max(x, 0) \, .
\end{align*}
\end{example}

\begin{example}
Let $\mS_i = \Diag((\sigma_j)_{j \in S_i})$. The proximal operator of the convex conjugate of the PET distance~\eqref{EQU:PET:CVX} can be computed element-wise as $[\Prox^{\mS_i}_{D_i^\ast}(y)]_j = \Prox^{\sigma_j}_{\varphi_j^\ast}(y_j)$. For each element, the proximal operator is given by
\begin{align*}
    \Prox^\sigma_{\varphi^\ast}(y) = \frac12 \left[w + 1 - \Bigl( (w - 1)^2 + 4\sigma b\Bigr)^{1/2} \right] \, ,
\end{align*}
where we again omitted the indices $j$ for readability and denoted $w = y + \sigma r$.
\end{example}

\section{Algorithm} \label{SEC:ALGORITHM}%
The saddle point problem~\eqref{EQU:SADDLE} (and therefore the PET reconstruction problem~\eqref{EQU:PROB:OPT:GENERIC}) can be solved with the PDHG~\cite{Chambolle2011}, see Algorithm~\ref{ALG:PDHG}. It consists of very simple operations involving only basic linear algebra, matrix-vector multiplications and the evaluations of proximal operators. As seen in line 4 of the pseudo-code, PDHG updates all dual variables simultaneously. Therefore, in line 4 and 5, the projection and backprojection that corresponds to the whole data set have to be evaluated. The idea of SPDHG, Algorithm~\ref{ALG:SPDHG}, is to only select one dual variable randomly in each iteration (line 4) and to perform the update accordingly (line 5 and 6). An important detail is the extrapolation in line 8 with the inverse of the probability $p_i$ that $i$ will be selected in each iteration. This guarantees the convergence as proven in Theorem~\ref{THE:ALG} below.

\subsection{Convergence}
SPDHG is guaranteed to converge for any $f_i$ and $g$ which are convex, proper and lsc. We now state a very general convergence result which can be derived from~\cite[Theorem 4.3]{Chambolle2017}. The actual proof is omitted here for brevity. For more details on convergence and convergence rates we refer the reader to \cite{Chambolle2017}.

\begin{theorem}[Convergence] \label{THE:ALG}
Assume that the sampling is proper, \ie~the probability $p_i$ for an index $i \in [n]$ to be sampled is positive. Let the step length parameters $\mT = \min_{i \in [n]} \mT_i, \mS_i$ be chosen such that for all $i \in [n]$ the following bound on the operator norm
\begin{align}
    \left\|\mS^{1/2}_i \mA_i \mT^{1/2}_i\right\|^2 < p_i  \label{EQU:STEPSIZE}
\end{align}
holds. Then for any initialization, the iterates $(x, y)$ of SPDHG (Algorithm \ref{ALG:SPDHG}) converge to a saddle point of \eqref{EQU:SADDLE} almost surely in a Bregman distance.
\end{theorem}

\begin{algorithm}[t]
\caption{Primal-Dual Hybrid Gradient~(PDHG) to solve~\eqref{EQU:SADDLE}. Default values given in brackets.} \label{ALG:PDHG}
\textbf{Input:}
iterates $x (= 0)$, $y (= 0)$,
step parameters $\mS = \{\mS_i\}$, $\mT$\\[-5mm]
\begin{algorithmic}[1]
    \State $\overline{z} = z = \mA^T y \; (= 0)$
    \For{$k = 1, \ldots$}
    \State $x = \Prox^\mT_g\left(x - \mT \overline{z}\right)$
    \State $y_i^+ = \Prox^{\mS_i}_{f^\ast_i}\left(y_i + \mS_i \mA_i x\right) \quad \text{for $i = 1, \ldots, n$}$
    \State $\Delta z = \sum_{i=1}^n \mA_i^T \left(y_i^+ - y_i\right)$
    \State $z = z + \Delta z, \quad y = y^+$
    \State $\overline{z} = z + \Delta z$
    \EndFor
\end{algorithmic}%
\end{algorithm}

\begin{algorithm}[t]
\caption{Stochastic Primal-Dual Hybrid Gradient~(SPDHG) to solve~\eqref{EQU:SADDLE}. Default values given in brackets.} \label{ALG:SPDHG}
\textbf{Input:}
iterates $x (= 0)$, $y (= 0)$,
step parameters $\mS = \{\mS_i\}$, $\mT$\\[-5mm]
\begin{algorithmic}[1]
    \State $\overline{z} = z = \mA^T y \; (= 0)$
    \For{$k = 1, \ldots$}
    \State $x = \Prox^\mT_g\left(x - \mT \overline{z}\right)$
    \State Select $i \in [n]$ at random with probability $p_i$
    \State $y_i^+ = \Prox^{\mS_i}_{f^\ast_i}\left(y_i + \mS_i \mA_i x\right)$
    \State $\Delta z = \mA_i^T \left(y_i^+ - y_i\right)$
    \State $z = z + \Delta z, \quad y_i = y_i^+$
    \State $\overline{z} = z + \frac{1}{p_i} \Delta z$
    \EndFor
\end{algorithmic}%
\end{algorithm}

\begin{remark}[Computational Efficiency]
Each iteration of Algorithm~\ref{ALG:SPDHG} is computationally efficient as only projections and backprojections corresponding to the randomly selected subset $i$ of the data are required. However, the algorithm maintains the whole backprojected dual variable $z = \mP^T y = \sum_{i=1}^m \mP_i^T y_i$ and in each iteration updates the primal variable with it.
\end{remark}

\begin{remark}[Memory Requirements]
The memory requirement of Algorithm~\ref{ALG:SPDHG} is higher compared to OSEM or gradient descent but still reasonably low. It requires memory equivalent to two images $(z, \overline{z})$ and up to twice the binned sinogram data ($y, y^+$) in addition to the necessary memory consumption (output image, sinogram data, background and normalization).
\end{remark}

\begin{remark}[Sampling]
SPDHG allows any kind of random selection as long as the draws are independent and the probability that block $i$ is being selected with positive probability $p_i > 0$. We will investigate two choices of sampling in the numerical section of this paper. A more thorough numerical and theoretical investigation will be subject of future work.
\end{remark}

\subsection{Step Sizes and Preconditioning}
We will now discuss two different choices of step sizes under which SPDHG is guaranteed to converge. The proof of the following theorem uses arguments from~\cite{Chambolle2017} and~\cite{Pock2011} and is omitted here for brevity.

\begin{theorem}[Step Size Parameters] \label{THE:STEPPARM}
Let $\rho < 1$ and $\gamma > 0$. Then, condition~\eqref{EQU:STEPSIZE} of Theorem~\ref{THE:ALG} is satisfied by
\begin{align}
   \mS_i = \gamma \frac{\rho}{\|\mA_i\|} \mI \; , \quad \mT_i = \gamma^{-1} \frac{\rho p_i}{ \|\mA_i\|} \mI \, . \label{EQU:STEPSIZE:SCALAR}
\end{align}
Moreover, if $\mA_i$ has only non-negative elements, then condition~\eqref{EQU:STEPSIZE} is also satisfied by
\begin{align}
   \mS_i = \gamma \Diag\left(\frac{\rho}{\mA_i 1}\right) \; , \quad \mT_i = \gamma^{-1}  \Diag\left(\frac{\rho p_i}{\mA_i^T 1}\right) \, . \label{EQU:STEPSIZE:PRECOND}
\end{align}
\end{theorem}

\begin{figure}%
\centering%
\begin{tikzpicture}%
\node[anchor=south west] at (0,0) {\includegraphics[height=3.4cm]{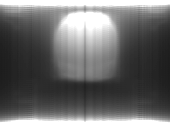}\hspace*{2mm}%
\includegraphics[height=3.4cm]{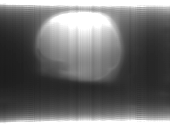}\hspace*{2mm}%
\includegraphics[height=3.4cm]{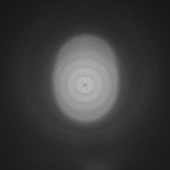}};%
\DrawLabel{Preconditioned step size $\mT$}
\end{tikzpicture}\\[-1mm]%
\begin{tikzpicture}%
\node[anchor=south west] at (0,0) {\includegraphics[height=3.2cm]{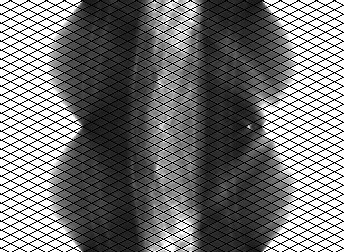}%
\includegraphics[height=3.2cm]{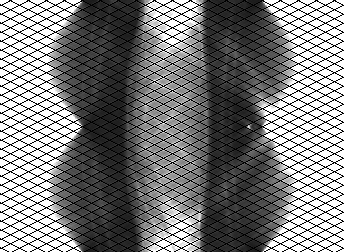}%
\includegraphics[height=3.2cm]{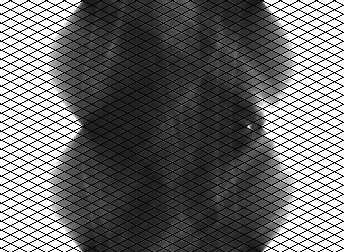}};%
\DrawLabel{Preconditioned step size $\mS$}
\end{tikzpicture}\\[-3mm]%
\caption{\textbf{Preconditioned parameters} $\mT$ (top) and $\mS$ (bottom)~\eqref{EQU:STEPSIZE:PRECOND} for the data set \dataFdg~(see section~\ref{section:data}). Apart from the boundary the step sizes are large in interesting regions, clearly showing the head of the patient.} \label{FIG:PARAM}%
\end{figure}

An example of preconditioned step sizes~\eqref{EQU:STEPSIZE:PRECOND} is shown in Figure~\ref{FIG:PARAM} .

\begin{remark}
If $n = 1$ and $p_i = 1$, then the step sizes~\eqref{EQU:STEPSIZE:SCALAR} can be identified with the scalar step sizes $\sigma_i = \gamma \rho / \|\mA_i\|$ and $\tau = \gamma^{-1} \rho / \|\mA_i\|$ which are commonly chosen for PDHG.
\end{remark}

\begin{remark}
Note that the non-negativity condition holds for the PET projection operator (and any other ray tracing based operator). Moreover, the step size $\mT$ in~\eqref{EQU:STEPSIZE:PRECOND} resembles the sensitivities used in the update of MLEM~\eqref{EQU:MLEM} and OSEM~\eqref{EQU:OSEM}. In addition, a similar preconditioning is performed for the dual variable in the data space.
\end{remark}

\section{Numerical Results} \label{sec:numerics}
The numerical experiments use the open-source package ODL~\cite{Adler2017odl} which allows for efficient algorithm prototyping in Python. The projection and backprojections are computed with CUDA in single-precision through the open-source package NiftyPET~\cite{Markiewicz2017} which is accessible via Python. All results in this section were obtained by selecting subsets with equidistantly divided angles. We use in all numerical experiments the parameter $\gamma=1$. Fine-tuning of this parameter is left for future work. Moreover, all peak signal-to-noise (PSNR) or relative objective comparisons are performed by first computing an approximate minimizer $x^\ast$ by the deterministic PDHG using \changed{5,000}{thousands of} iterations. The PSNR is defined as $\operatorname{PSNR}(x^k, x^\ast) = 20 \log(\|x^\ast\|_\infty / \|x^k - x^\ast\|_2)$ and the relative objective value is defined as $(\Psi(x^k) - \Psi(x^\ast)) / (\Psi(x^0) - \Psi(x^\ast))$. We frequently use the word "epoch" to denote the number of iterations of a randomized algorithm which are in expectation computationally equivalent to one iteration of the deterministic algorithm that uses all data for each iteration. As an example, if a randomized algorithm only uses 1/10 of the data in each iteration, then after 10 iterations one can expect that the algorithm has used all data, thus in this case 1 epoch equals 10 iterations.
In all figures, the dashed lines correspond to deterministic and the solid lines to randomized algorithms. {\color{blue}The Python code and one data set will be made accessible upon acceptance of this manuscript.}

\subsection{Data} \label{section:data}
We validate the numerical performance of the proposed algorithm on two clinical PET data sets which we refer to as \dataFdg~and \dataAmy. The two separate PET brain datasets each use a distinct radiotracer: [$^{18}$F]FDG for epilepsy and [$^{18}$F]florbetapir for the neuroscience sub-study Insight'46 of the Medical Research Council National Survey of Health and Development~\cite{Lane2017}. The epileptic patient was injected with 250 Mbq of FDG, one hour before the 15-minute PET acquisition.  The neuroscience volunteer was injected with 370 MBq of florbetapir and scanned dynamically for one hour, starting at the injection time.  The last ten minutes were used as a measurement of amyloid deposition, which for the participant was negative.

\subsection{Results for Total Variation}
In this section we analyze the impact of various choices within SPDHG on its performance, from randomness over sampling to preconditioning. The test case is total variation prior as defined in~\eqref{EQU:SETTING:TV}.

\subsubsection{Randomness} 
\begin{figure}[h!]%
\InclFig{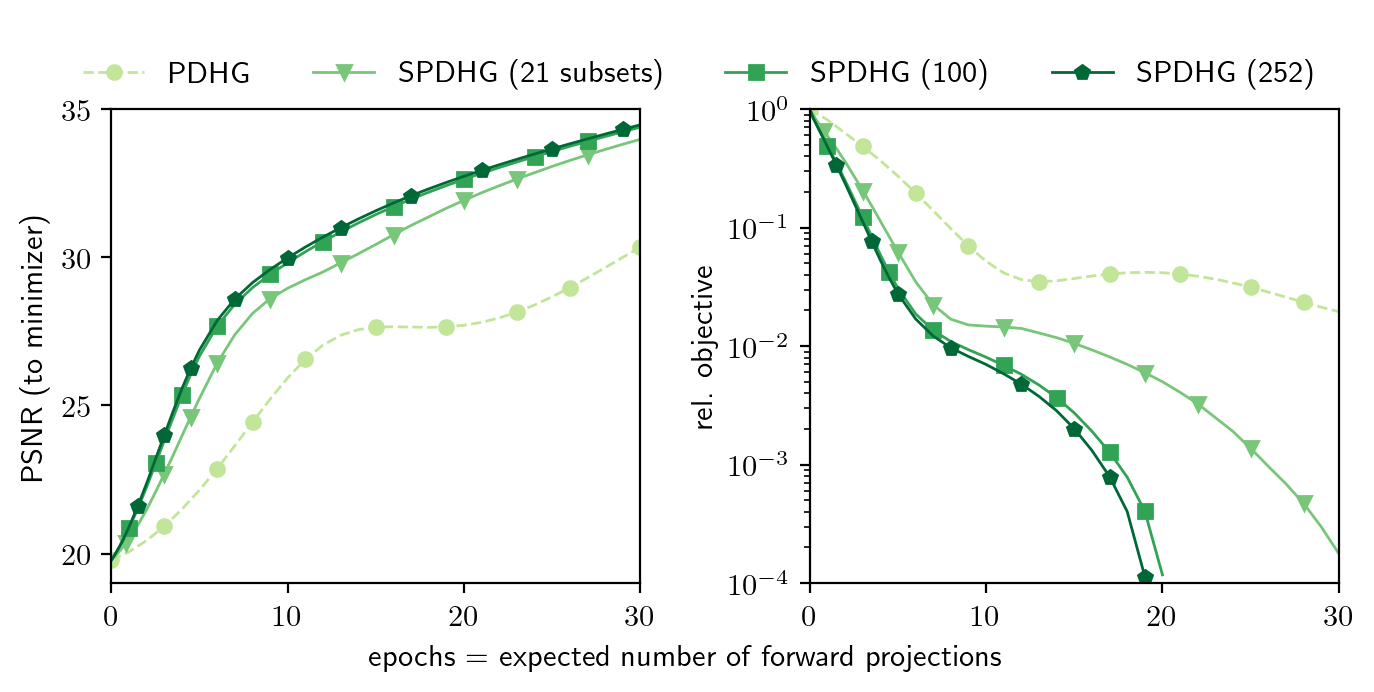}%
\caption{\textbf{Deterministic v randomized.} The results for the data set \dataFdg~with TV prior show that the randomized algorithms are much faster than their deterministic counterpart. Moreover, more subsets leads to a faster algorithm.} \label{FIG:TV:RANDOM}
\end{figure}

Figure~\ref{FIG:TV:RANDOM} shows the effect of randomness where we compare the deterministic PDHG to SPDHG with uniform sampling and scalar step sizes~\eqref{EQU:STEPSIZE:SCALAR} for two different number of subsets. The horizontal axis reflects the number of projections in each algorithm, we call one full projection for the whole data one ``epoch''. Here and in the following dashed lines represent deterministic and solid lines randomized algorithms. We can easily see that both random variants are faster than then deterministic PDHG. Moreover, the randomized SPDHG becomes faster by choosing a larger number of subsets.

\subsubsection{Sampling} 
\begin{figure}[h!]%
\InclFig{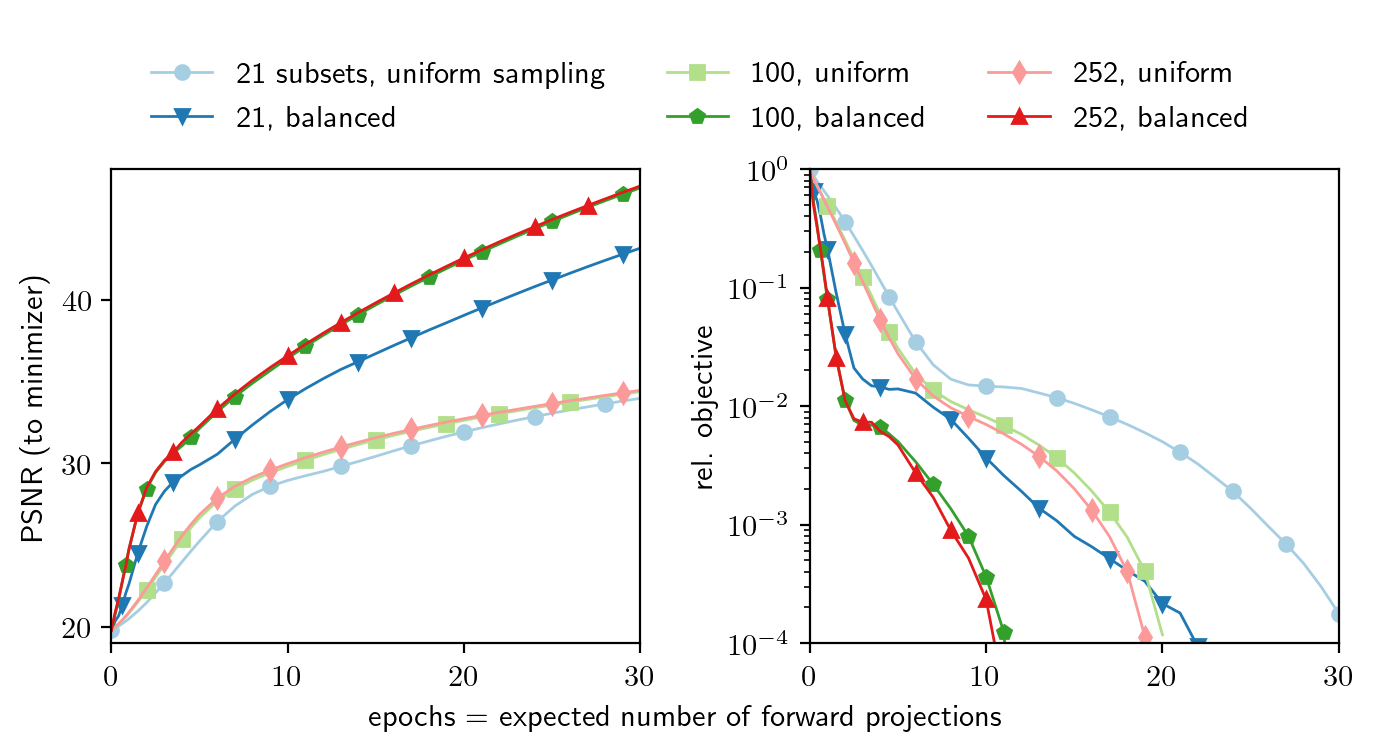}%
\caption{\textbf{Uniform v balanced sampling.} In addition to increasing the number of subsets, the sampling is also very important for the speed of the algorithm: 21 subsets with balanced sampling is faster than 100 subsets with uniform sampling.} \label{FIG:TV:SAMPLING}
\end{figure}

The effect of different choices of sampling is shown in Figure~\ref{FIG:TV:SAMPLING}. We compare two different samplings: uniform sampling and balanced sampling. The uniform sampling chooses all indices $i \in [n]$ with equal probability $p_i = 1/n$. In contrast, for balanced sampling we choose with uniform probability either data or prior. If we choose data, then we select a subset again randomly with uniform probability. Thus, the probability for each subset of the data to be selected is $p_i = 1 / (2m), i \in [m]$ and for the prior to be selected $p_n = 1/2$.

We make two observations. First, balanced sampling is always faster than uniform sampling. This shows the importance of updating the dual variable associated to the prior. Second, for either sampling choosing a larger number of subsets again improves the performance.

\subsubsection{Preconditioning} 
\begin{figure}[h!]%
\InclFig{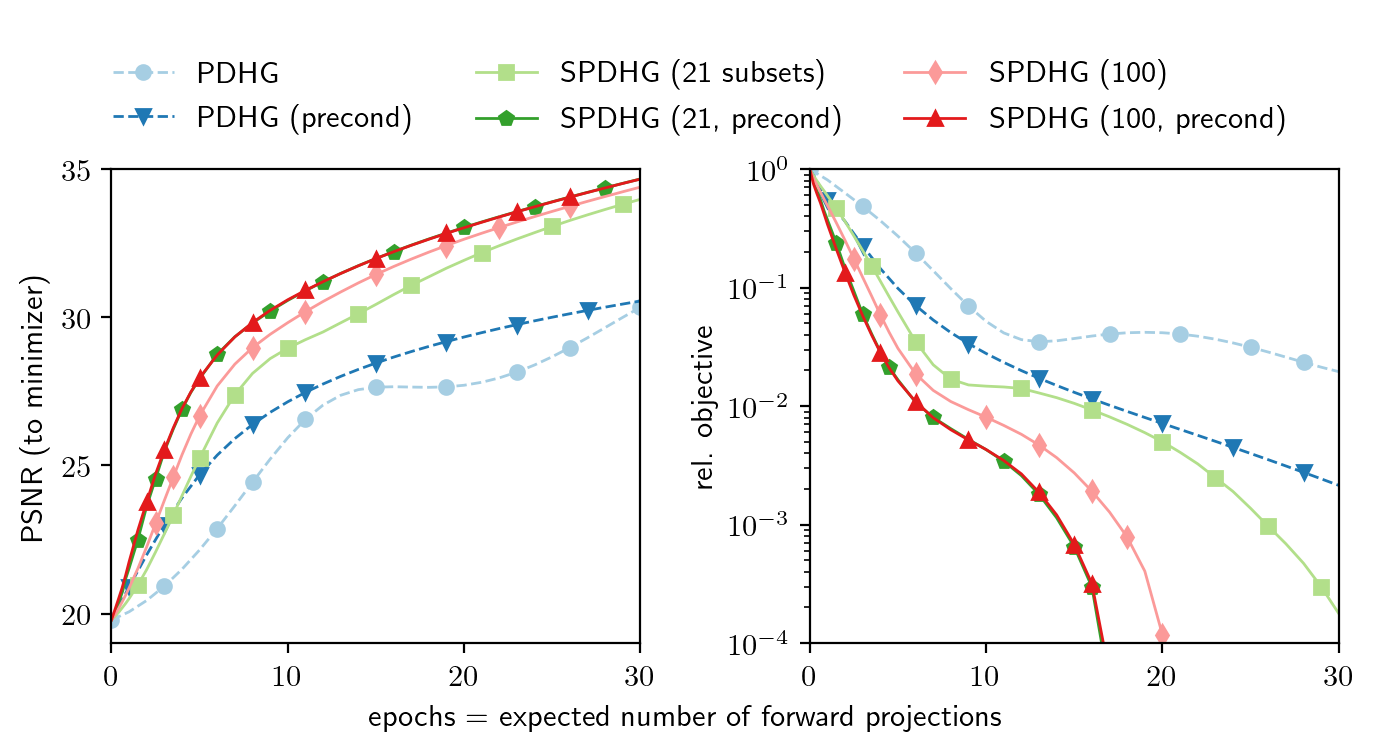}%
\caption{\textbf{Preconditioning} can be used with and without randomization. The preconditioned algorithms are much faster than without preconditioning.} \label{FIG:TV:PRECON}%
\end{figure}

As shown in Theorem~\ref{THE:STEPPARM}, the step size parameters $\mT$ and $\mS_i$ can be chosen either as scalars~\eqref{EQU:STEPSIZE:SCALAR} or as vectors~\eqref{EQU:STEPSIZE:PRECOND}, the latter can be seen as a form of preconditioning. Results are shown in Figure~\ref{FIG:TV:PRECON}, where we see that preconditioning may accelerate the convergence of either the deterministic PDHG or the randomized SPDHG. Moreover, combining randomization and preconditioning yields an even faster algorithm.

\subsubsection{Performance of Proposed Algorithm} 
\begin{figure}[h!]
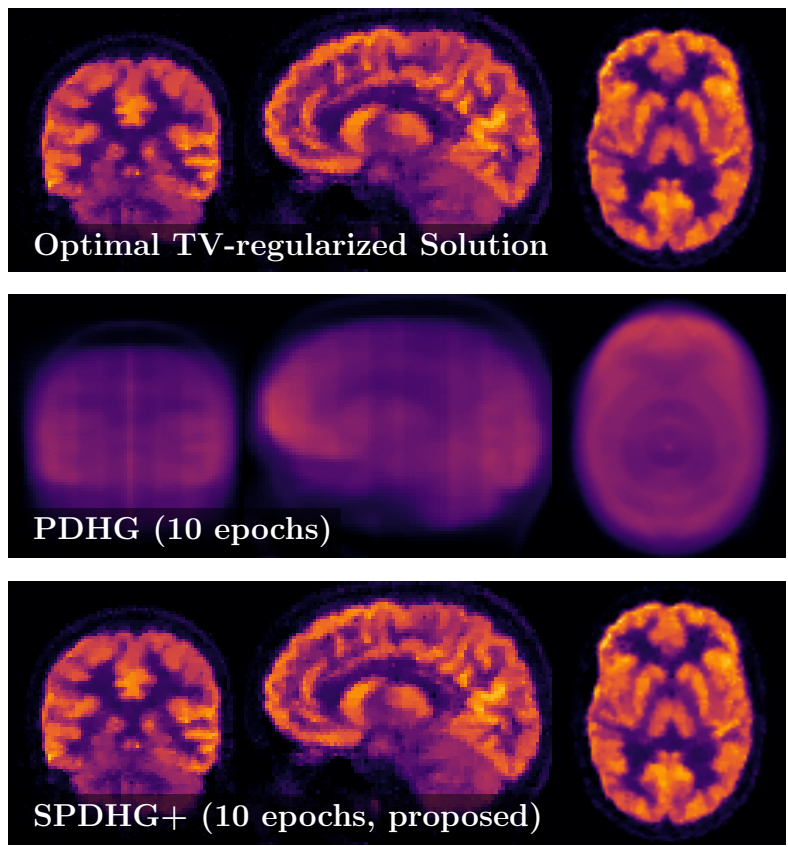
%
\centering
\PlotFdg{Optimal TV-regularized Solution}{map_tv}{inferno}{target}\\%
\PlotFdg{PDHG (10 epochs)}{map_tv}{inferno}{PDHG1_10}\\%
\PlotFdg{SPDHG+ (10 epochs, proposed)}{map_tv}{inferno}{SPDHG2-252-bal_10}%
\caption{\textbf{Qualitative results} show that in contrast to the deterministic PDHG, the proposed SPDHG+ (252 subsets) approximates the optimal solution well after only 10 epochs. \changed{The "optimal" solution was computed with 5,000 iterations of PDHG.}{}} \label{FIG:FDG:TV}%
\end{figure}

Based on the previous three examples, we propose to combine randomization, balanced sampling and preconditioning, which we refer to as SPDHG+. Figure~\ref{FIG:FDG:TV} shows the visual performance of PDHG and SPDHG+. In contrast to the deterministic PDHG, the proposed SPDHG+ yields a good approximation of the optimal solution after only 10 epochs.

\subsection{Further Numerical Results}
\subsubsection{Anisotropic Total Variation}
\begin{figure}[h!]%
\centering%
\PlotAmy{Optimal Solution}{map_atv}{inferno}{target}\\%
\PlotAmy{PDHG (10 epochs)}{map_atv}{inferno}{PDHG1_10}\\%
\PlotAmy{SPDHG+ (10 epochs, proposed)}{map_atv}{inferno}{SPDHG2-252-bal_10}%
\vspace*{4mm}
\InclFig{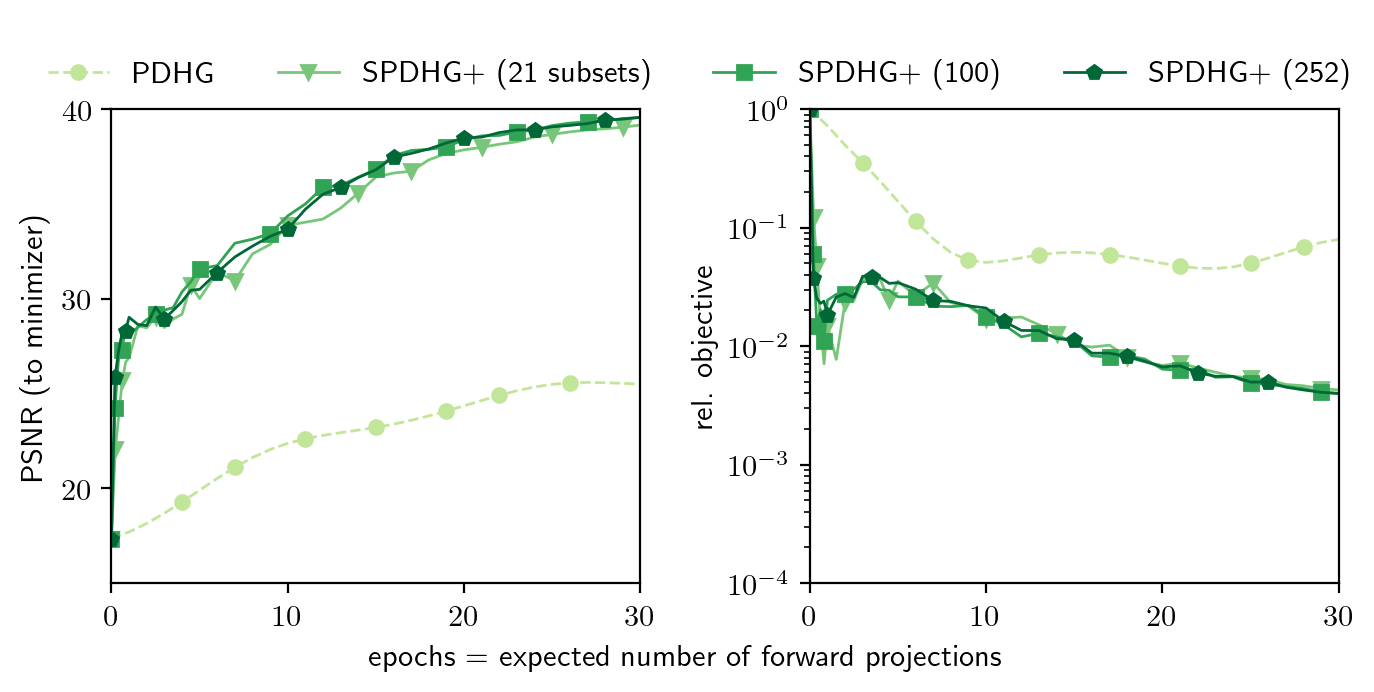}%
\caption{\textbf{Anisotropic TV} regularized reconstruction from \dataFdg~data. Top: PDHG and SPDHG+ (252 subsets) reconstructions after 10 epochs. Bottom: Quantitative results show a significant speed-up from randomization and preconditioning. Increasing the number of subsets from 21 to 252 has little effect on this data set. \changed{The "optimal" solution was computed with 5,000 iterations of PDHG.}{}} \label{FIG:AMY:ATV}%
\end{figure}

Anisotropic total variation decouples the penalization of the derivatives. The mathematical model is similar to the isotropic TV model~\eqref{EQU:SETTING:TV}, the only difference being the norm how the total variation is measured: $f_n = \alpha \|\cdot\|_{1,1}$. It can be seen in Figure~\ref{FIG:AMY:ATV} for \dataAmy~that with randomization and preconditioning only a few epochs are needed to obtain a good approximation of the optimal solution.

\subsubsection{Directional Total Variation}
\begin{figure}[h!]%
\centering%
\PlotAmy{Optimal Solution}{map_dtv}{inferno}{target}\\%
\PlotAmy{PDHG (10 epochs)}{map_dtv}{inferno}{PDHG1_10}\\%
\PlotAmy{SPDHG+ (10 epochs, proposed)}{map_dtv}{inferno}{SPDHG2-252-bal_10}%
\vspace*{4mm}
\InclFig{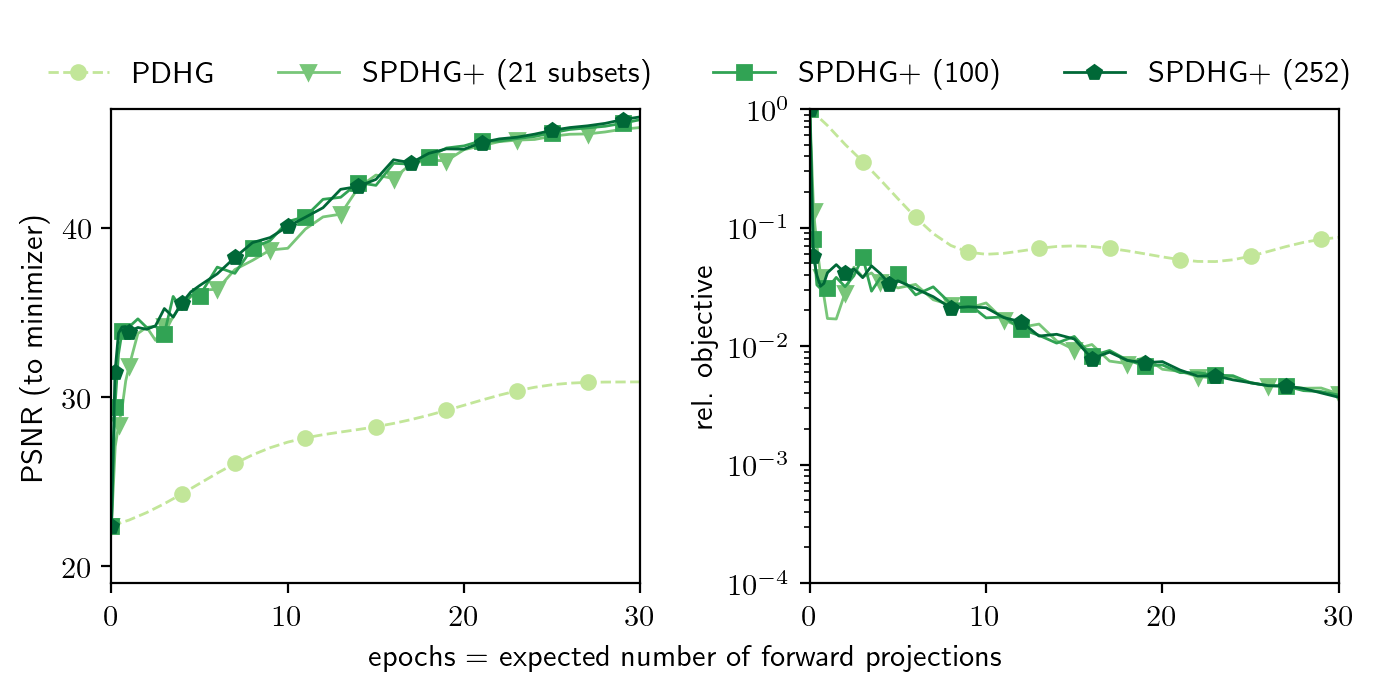}%
\caption{\textbf{Directional TV} prior (which uses MRI information) for \dataAmy~data. Both qualitative (top) and quantitative results (bottom) show the speed up provided by randomization and preconditioning. \changed{The "optimal" solution was computed with 5,000 iterations of PDHG.}{}} \label{FIG:AMY:DTV}%\vspace*{-4mm}%
\end{figure}

Anatomical information from a co-registered MRI is available on combined PET-MR scanners. The structural information of the anatomy can be utilized by the directional total variation prior, see~\cite{Ehrhardt2016a, Ehrhardt2016b, Schramm2017petplusmri, Bungert2018remotesensing} for details. The mathematical model is similar to the total variation model~\eqref{EQU:SETTING:TV}, except for an additional matrix $\mD$. Thus, the only difference is $\mA_n = \mD \nabla$. A numerical example is shown in Figure~\ref{FIG:AMY:DTV} for the data set \dataAmy.

\subsubsection{Total Generalized Variation}
More sophisticated regularization can be achieved by the total generalized variation (TGV)~\cite{Bredies2010, Bredies2015tgvnumerics}%
\begin{align*}
\TGV_{\alpha_0, \alpha_1}(u) = \inf_w \left\{ \alpha_0 \|\nabla u - w\|_{2,1} + \alpha_1 \|\mathcal E w\|_{2,1} \right\}
\end{align*}
which can balance first and second order regularization and achieves edge-preserved reconstruction while avoiding the stair-casing artifact. We can solve the TGV regularized PET reconstruction problem by solving problem~\eqref{EQU:PROB:OPT:GENERIC} with the assignment $x = (u, w)$ and
\begin{align*}
\begin{aligned}
    n &= m + 2, & \mA_i &= (\mP_i, 0), i \in [m], & \mA_{n-1} &= (\nabla, - \mI), & \mA_n &= (0, \mathcal E)\\
    g(x) &= \imath_+(u), & f_i &= D_i, i \in [m], & f_{n-1} &= \alpha_0 \|\cdot\|_{2,1}, & f_n &= \alpha_1 \|\cdot\|_{2,1} \, ,
\end{aligned} \label{EQU:SETTING:TGV}
\end{align*}
where $\mathcal E$ is a symmetrized gradient operator, see~\cite{Bredies2010, Bredies2015tgvnumerics} for more details.

\begin{figure}[h!]%
\centering
\PlotFdgZY{Optimal Solution: $u$}{map_tgv}{inferno}{target}\hspace*{-3mm}%
\PlotFdgZY{$w$}{map_tgv}{gray}{target_norm_vfield}\\%
\PlotFdgZY{PDHG (10 epochs): $u$}{map_tgv}{inferno}{PDHG1_10}\hspace*{-3mm}%
\PlotFdgZY{$w$}{map_tgv}{gray}{PDHG1_10_norm_vfield}\\%
\PlotFdgZY{SPDHG+$^*$ (10 epochs): $u$}{map_tgv}{inferno}{SPDHG2-252-bal_10}\hspace*{-3mm}%
\PlotFdgZY{$w$}{map_tgv}{gray}{SPDHG2-252-bal_10_norm_vfield}%
\vspace*{4mm}
\InclFig{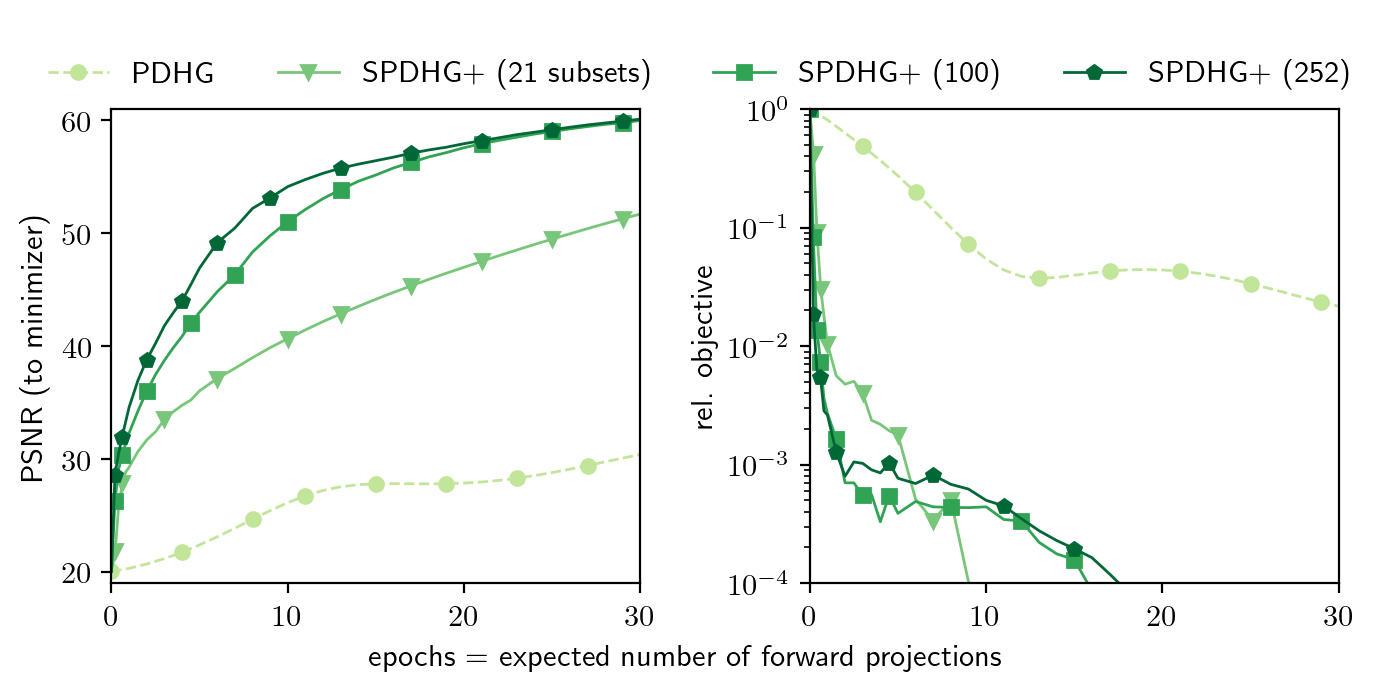}%
\caption{\textbf{TGV} regularized reconstruction for the \dataFdg~data. Only a few epochs are needed to approximate the optimal solution with randomization and preconditioning. This is visible for both the actual images $u$ and for the reconstructed vector field $w$. \changed{The "optimal" solution was computed with 5,000 iterations of PDHG.}{} $^*$proposed} \label{FIG:FDG:TGV}%
\end{figure}

The numerical results shown in Figure~\ref{FIG:FDG:TGV} are in line with the previous findings indicating that randomization and preconditioning can significantly speed up the reconstruction. However, we notice a significant increase in performance by increasing the number of subsets from 21 to 252.

\subsubsection{Comparison of Mathematical Models}
We conclude this section by a comparison of various methods on both data sets in Figures~\ref{FIG:FDG:METHODS} and~\ref{FIG:AMY:METHODS}. While we leave the detailed visual comparisons to the reader, we would like to note that all these images use the same number of projections so have basically the same computational cost.

\begin{figure}%
\centering%
\PlotFdgZY{ML}{ml}{inferno}{smoothed_SPDHG2-252_10}%
\PlotFdgZY{TV}{map_tv}{inferno}{SPDHG2-252-bal_10}\\%
\PlotFdgZY{aTV}{map_atv}{inferno}{SPDHG2-252-bal_10}%
\PlotFdgZY{TGV}{map_tgv}{inferno}{SPDHG2-252-bal_10}\\%
\PlotFdgZY{dTV (using MRI)}{map_dtv}{inferno}{SPDHG2-252-bal_10}%
\PlotFdgZY{MRI structure for dTV}{map_dtv}{gray}{image_norm_sideinfo_grad}\\%
\caption{Comparison of several reconstruction approaches for the \dataFdg~data. All approaches have about the same computational cost (10 epochs).} \label{FIG:FDG:METHODS}%
\end{figure}%
\begin{figure}%
\centering%
\PlotAmyZY{ML}{ml}{inferno}{smoothed_SPDHG2-252_10}%
\PlotAmyZY{TV}{map_tv}{inferno}{SPDHG2-252-bal_10}\\%
\PlotAmyZY{aTV}{map_atv}{inferno}{SPDHG2-252-bal_10}%
\PlotAmyZY{TGV}{map_tgv}{inferno}{SPDHG2-252-bal_10}\\%
\PlotAmyZY{dTV (using MRI)}{map_dtv}{inferno}{SPDHG2-252-bal_10}%
\PlotAmyZY{MRI structure for dTV}{map_dtv}{gray}{image_norm_sideinfo_grad}\\%
\caption{Comparison of several reconstruction approaches for the \dataAmy~data. All approaches have about the same computational cost (10 epochs).} \label{FIG:AMY:METHODS}%
\end{figure}

\section{Discussion}%
The extensive numerical experiments all consistently confirm that randomization and preconditioning both speed up the reconstruction. These trends were irrespective of the data set and the chosen prior. The convergence speed in our work was abstractly defined by a solution of the underlying mathematical optimization model approximated with way too many iterations than would be feasible in routine clinical practice. This strategy was chosen intentionally as we did not want to target a specific clinical use case. After these successful initial trials, in the future we will collaborate with medical researchers and clinicians to focus on specific use cases where each use case defines its own metric of what images we wish to reconstruct.

The focus of this contribution was on non-smooth priors like total variation and its descendants like total generalized variation and directional total variation. However, as long as the proximal operators are simple to evaluate, the proposed randomized and preconditioned algorithm can be applied to any other model, too. It would be of interest to compare this algorithm to convergent subset accelerated algorithms for smooth priors like BSREM~\cite{DePierro2001, Ahn2003}, TRIOT~\cite{Ahn2006} and OS-SPS~\cite{Ahn2003}.

We highlighted the improvements from choosing different distributions for subset selection by comparing ``uniform'' and ``balanced sampling''. Further improvements are expected by optimizing the probability selection of this algorithm. This can either be an optimal distribution that is constant along the iterations or even developing over the course of the iterations. We will investigate this direction further in the future.

With the exception of Figures~\ref{FIG:AMY:ATV} and~\ref{FIG:AMY:DTV} where 21, 100 and 252 subsets were similarly fast, more subsets always resulted in a faster algorithm. There are neither theoretical nor numerical insights how the speed will depend on the subset selection and if more subsets always result in a faster algorithm. However, the numerical evidence suggests that increasing the number of subsets never decreases the speed of the algorithm. This being said, due to the per iteration computational costs, from a practical point of view, there will be an optimal number of subsets that might depend on the prior and even the data (\eg~number of counts) to be reconstructed. We would like to point out that the two Figures~\ref{FIG:AMY:ATV} and~\ref{FIG:AMY:DTV} have in common that both used the same tracer \dataAmy. In future work we will study the tracer-dependence of the convergence speed in more detail.

Moreover, the algorithm does not exploit any special structure of our optimization problem like smoothness or strong convexity. It is likely that exploiting these properties will lead to additional speed-up. However, as these properties for the PET data term depend on the acquired data, it is unlikely that a straightforward approach will be sufficient and a tailored solution will be necessary.

\section{Conclusion}%
We introduced a convergent subset accelerated algorithm for the reconstruction of PET images with non-smooth priors. The algorithm was enhanced by data-dependent preconditioning. Our numerical results showed that using both randomized subset selection and preconditioning can dramatically speed up the convergence of an iterative reconstruction algorithm. It was observed that a computational effort similar to the current clinical standard OSEM was sufficient for many non-smooth priors, showing that these are now, for the first time, feasible to be used in daily clinical routine.

While these observations were consistent among two data sets with different tracers, more studies are needed to confirm the benefits of this reconstruction strategy. Overall, this algorithmic advancement has the potential to \changed{change}{hugely impact} the PET reconstruction landscape as advanced mathematical models can now be combined with efficient and convergent subset acceleration.%

\ack%
M.J.E. and C.-B.S. acknowledge support from Leverhulme Trust project ``Breaking the non-convexity barrier'', EPSRC grant ``EP/M00483X/1'', EPSRC centre ``EP/N014588/1'', the Cantab Capital Institute for the Mathematics of Information, and from CHiPS and NoMADS (Horizon 2020 RISE project grants). In addition, M.J.E. acknowledges support from the EPSRC platform grant ``EP/M020533/1''. Moreover, C.-B.S. is thankful for support by the Alan Turing Institute. In addition, all authors gratefully acknowledge the hardware donation by the NVIDIA Corporation.

\bibliographystyle{ieeetr}
\bibliography{library}

\end{document}